\documentclass{PoS}
\usepackage{subfigure}
\usepackage{url}

\def\BBz{0$\nu\beta\beta$}

\def\mj{M{\sc ajo\-ra\-na}}
\def\dem{D{\sc emonstrator}}

\def\QBB{Q$_{\beta\beta}$}

\def\ge{$^{76}$Ge}

\title{The \mj\ \dem\ search for neutrinoless double beta decay}

\ShortTitle{The \mj\ \dem\ search for neutrinoless double beta decay}

\newcommand{\blhill}{Department of Physics, Black Hills State University, Spearfish, SD, USA}
\newcommand{\ITEP}{National Research Center ``Kurchatov Institute'' Institute for Theoretical and Experimental Physics, Moscow, Russia}
\newcommand{\JINR}{Joint Institute for Nuclear Research, Dubna, Russia}
\newcommand{\lbnl}{Nuclear Science Division, Lawrence Berkeley National Laboratory, Berkeley, CA, USA}
\newcommand{\lanl}{Los Alamos National Laboratory, Los Alamos, NM, USA}
\newcommand{\queens}{Department of Physics, Engineering Physics and Astronomy, Queen's University, Kingston, ON, Canada}
\newcommand{\uw}{Center for Experimental Nuclear Physics and Astrophysics,
and Department of Physics, University of Washington, Seattle, WA, USA}
\newcommand{\unc}{Department of Physics and Astronomy, University of North Carolina, Chapel Hill, NC, USA}
\newcommand{\duke}{Department of Physics, Duke University, Durham, NC, USA}
\newcommand{\ncsu}{Department of Physics, North Carolina State University, Raleigh, NC, USA}	
\newcommand{\ornl}{Oak Ridge National Laboratory, Oak Ridge, TN, USA}
\newcommand{\ou}{Research Center for Nuclear Physics, Osaka University, Ibaraki, Osaka, Japan}
\newcommand{\pnnl}{Pacific Northwest National Laboratory, Richland, WA, USA}
\newcommand{\princeton}{Department of Physics, Princeton University, Princeton, NJ, USA}
\newcommand{\ttu}{Tennessee Tech University, Cookeville, TN, USA}
\newcommand{\sdsmt}{South Dakota School of Mines and Technology, Rapid City, SD, USA}
\newcommand{\usc}{Department of Physics and Astronomy, University of South Carolina, Columbia, SC, USA}
\newcommand{\usd}{Department of Physics, University of South Dakota, Vermillion, SD, USA}
\newcommand{\ut}{Department of Physics and Astronomy, University of Tennessee, Knoxville, TN, USA}
\newcommand{\tunl}{Triangle Universities Nuclear Laboratory, Durham, NC, USA}

\author{\speaker{C.~Cuesta}, M.~Buuck, J.A.~Detwiler, J.~Gruszko, I.S.~Guinn, J.~Leon, and R.G.H.~Robertson \\
        \uw \\
        E-mail: \email{ccuesta@uw.edu}}

\author{N. Abgrall, A.W. Bradley, Y-D. Chan, S. Mertens, A.W.P. Poon, and K. Vetter\thanks{Alternate address: Department of Nuclear Engineering, University of California, Berkeley, CA, USA} \\
		\lbnl}

\author{I.J. Arnquist, E.W. Hoppe, R.T. Kouzes, and J.L. Orrell \\
		\pnnl}

\author{F.T.~Avignone~III \\
        \usc \\ \ornl}
			
\author{A.S. Barabash, S.I. Konovalov, and V. Yumatov \\
		\ITEP}

\author{F.E. Bertrand, A. Galindo-Uribarri, D.C. Radford, R.L. Varner, and C.-H. Yu \\
		\ornl}

\author{V. Brudanin, M. Shirchenko, S. Vasilyev, E. Yakushev, and I. Zhitnikov \\
		\JINR}

\author{M. Busch   \\
		\duke \\ \tunl}

\author{T.S.~Caldwell, T.~Gilliss, R.~Henning, M.A.~Howe, J.~MacMullin, S.J.~Meijer,
C.~O'Shaughnessy, J.~Rager, B.~Shanks, J.E.~Trimble, K.~Vorren, and W.~Xu \thanks{Current address:  Department of Physics, University of South Dakota, Vermillion, SD, USA}\\
		\unc \\ \tunl}

\author{C.D. Christofferson, C. Dunagan, and A.M. Suriano \\
		\sdsmt}

\author{P.-H. Chu, S.R. Elliott, R. Massarczyk, K. Rielage, and B.R. White \\
		\lanl}

\author{ Yu. Efremenko and A.M. Lopez \\
		\ut}

\author{H.~Ejiri \\
		\ou}

\author{A. Fullmer \\
		\ncsu \\ \tunl}

\author{G.K.~Giovanetti \\
	    \princeton}

\author{M.P.~Green \\
		\ncsu \\ \tunl \\ \ornl}

\author{V.E. Guiseppe, D. Tedeschi, and C. Wiseman \\
		\usc}

\author{B.R.~Jasinski \\
		\usd}

\author{K.J.~Keeter \\
		\blhill}

\author{M.F.~Kidd \\
		\ttu}

\author{R.D.~Martin \\
		\queens}

\author{E. Romero-Romero\\
		\ut \\ \ornl}

\author{J.F. Wilkerson\\
		\unc \\ \tunl \\ \ornl}

\abstract{The \mj\ Collaboration is constructing the \mj\ \dem, an ultra-low background, modular, HPGe detector array with a mass of 44.8-kg (29.7~kg enriched $\geq$88\% in \ge) to search for neutrinoless double beta decay in $^{76}$Ge. The next generation of tonne-scale Ge-based neutrinoless double beta decay searches will probe the neutrino mass scale in the inverted-hierarchy region. The \mj\ \dem\ is envisioned to demonstrate a path forward to achieve a background rate at or below 1 count/tonne/year in the 4~keV region of interest around the Q-value of 2039~keV. The~\mj\ \dem\ follows a modular implementation to be easily scalable to the next generation experiment. First data taken with the \dem\ are introduced here.}

\FullConference{XIII International Conference on Heavy Quarks and Leptons\\
		22-27 May, 2016\\
		Blacksburg, Virginia, USA}

\begin{document}

\section{Introduction}
\label{sec1}

In a number of even-even nuclei, $\beta$ decay is energetically forbidden, but the second order weak process of 2$\nu$ double beta decay is allowed, as first proposed by Goeppert-Mayer in 1935~\cite{gmayer}. If the neutrino is a Majorana particle, neutrinoless double beta (\BBz) decay could also occur via the exchange of a light Majorana neutrino, or by other mechanisms~\cite{Avignone,vergados}. This decay violates lepton number and provides a model-independent test of the nature of the neutrino. The rate of \BBz-decay via light Majorana neutrino exchange is given by

\begin{equation}\label{eq1}
\left(T^{0\nu}_{1/2}\right)^{-1}=G^{0\nu}|M_{0\nu}|^{2} \left( \frac{\langle m_{\beta\beta}\rangle}{m_{e}} \right)^{2}
\end{equation}

\noindent where $G^{0\nu}$ is a phase space factor, $M_{0\nu}$ is a nuclear matrix element, and $m_{e}$ is the electron mass. $\langle m_{\beta\beta}\rangle$ is the effective Majorana neutrino mass of the exchanged neutrino. The latter is given by $\langle m_{\beta\beta}\rangle=\left|\sum_{i=1}^{3}U_{ei}^{2}m_{i}\right|$, where $U_{ei}$ specifies the admixture of neutrino mass eigenstate $i$ in the electron neutrino. Because $\langle m_{\beta\beta}\rangle$ depends on the oscillation parameters, both the overall neutrino mass and the mass hierarchy can impact the observed rate. Provided the nuclear matrix elements are are well evaluated, \BBz-decay experiments could establish an absolute scale for the neutrino mass.


Experimentally, \BBz-decay can be detected by searching the spectrum of the summed energy of the emitted betas for a monoenergetic line at the Q-value of the decay (\QBB). Recent sensitive searches for~\BBz\ carried out in $^{76}$Ge (GERDA~\cite{GERDA,GERDAII}), $^{136}$Xe (KamLAND-Zen~\cite{KamLAND,KamLANDII} and EXO-200~\cite{EXO,EXO2}), $^{130}$Te (CUORE-0~\cite{CUORE0}), among others, set limits on the decay half-life.

\section{Overview of the \mj~\dem}
\label{sec2}

The~\mj~\dem~\cite{mjd} is an array of enriched and natural germanium detectors that will search for the \BBz-decay of \ge. The specific goals of the~\mj~\dem~are several: to demonstrate a path forward to achieving a background rate at or below 1~count/(ROI-t-y) in the 4~keV region of interest (ROI) around the 2039~keV~\QBB~of the \ge\ 0\BBz-decay when scaled up to a tonne scale experiment; show technical and engineering scalability toward a tonne-scale instrument; and perform searches for other physics beyond the Standard Model, such as dark matter and axions.

The experiment is composed of 44.8~kg of high-purity Ge (HPGe) detectors which also act as the source of \ge\ \BBz-decay. The benefits of HPGe detectors are that Ge is an intrinsically low-background source material, with understood enrichment chemistry, excellent energy resolution, and event reconstruction capabilities. P-type point contact detectors~\cite{ppc,ppc2} were chosen after extensive R\&D by the collaboration for their powerful background rejection capabilities. The \dem\ consists of a mixture of HPGe detectors including, 29.7 kg built from Ge material that is enriched to $\geq$88\% in \ge\ and 15.1~kg fabricated from natural Ge (7.8\% \ge). The average mass of the enriched detectors is $\sim$850~g.

\begin {figure}[ht]
\includegraphics[width=0.65\textwidth]{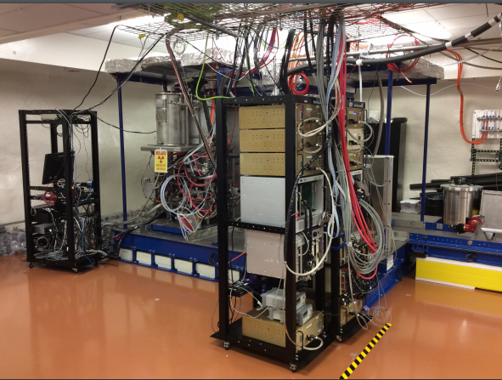}
\centering \caption{\it Picture of the \textsc{Majorana Demonstrator}.}
\label{fig:section}
\end {figure}

A modular instrument composed of two cryostats built from ultra-pure electroformed copper is being constructed. Each module hosts 7 strings of 3-5 detectors. The modules are operated in a passive shield that is surrounded by a 4$\pi$ active muon veto. To mitigate the effect of cosmic rays and prevent cosmogenic activation of detectors and materials, the experiment is being deployed at 4850~ft depth (4260~m.w.e. overburden) at the Sanford Underground Research Facility in Lead, SD~\cite{surf}. A picture of the \mj~\dem\ is shown in Figure~\ref{fig:section}.

The main technical challenge of the \mj\ \dem\ is to reach a background rate of 3~counts/(ROI-t-y) after analysis cuts, which projects to a background level of 1~count/(ROI-t-y) in a large scale experiment after accounting for additional improvements from thicker shielding, better self-shielding, and if necessary, increased depth. To achieve this goal, background sources must be reduced and offline background rejection must be maximized. The estimated ROI contributions based on achieved assays~\cite{mjdassay} of materials and simulations for cosmic muon interactions sum to $<$3.5~counts/(ROI-t-y) in the \mj\ \dem.

\section{The \mj~\dem\ implementation}
\label{sec3}

The~\mj~\dem~follows a modular implementation to scale easily to the next generation experiment. The modular approach allows the independent assembly and commissioning of each module independently, providing a fast deployment and minimizing interference with already-operational detectors.

As a first step, a prototype module was constructed using a commercial copper cryostat. It was loaded with three strings of natural-abundance germanium and placed into shielding. Data was collected with this module from June 2014 through July 2015. It served as a test bench for mechanical designs, fabrication methods, and assembly procedures for the construction of the electroformed-copper Modules~1~\&~2. In addition, the prototype also tested DAQ, data building and analysis tools.

\begin {figure}[ht]
\includegraphics[height=0.27\textheight]{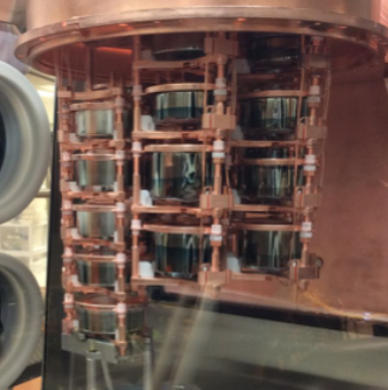}
\centering \caption{\it Module~2 detectors being installed.}
\label{fig:string}
\end {figure}

Following the prototype run, construction began on two modules with electroformed copper cryostats.  The first, Module~1, was assembled in 2015. Module~1 houses 16.8~kg of enriched germanium detectors and 5.7~kg of natural germanium detectors. The strings were assembled and characterized in dedicated String Test Cryostats. Module~1 was moved into the shield, and data taking began, during 2015. The final stage, Module~2 supports 12.8~kg of enriched and 9.4~kg of natural Ge detectors. Module~2 has been assembled and will start taking data soon. Figure~\ref{fig:string} shows the installation of detectors into the cryostat of Module~2.

\section{Module 1 preliminary results}
\label{sec4}

All open data from Module 1 are divided in two datasets: dataset~0 (DS0) and dataset~1 (DS1). DS0 was a set of commissioning runs used to test analysis and data production corresponding to data taken from June 2015 to October 2015. In fall 2015, we implemented planned improvements:
\begin{itemize}
\item Installed the electroformed inner copper shield. Built with the final pieces of electroformed copper, the inner shield was not yet ready for installation during the initial construction of Module~1.
\item Added additional shielding within the vacuum of the cross arm.
\item Replaced the cryostat Kalrez seal with PTFE which has much better radiopurity and much lower mass and entails a 3 orders of magnitude reduction in the ROI contribution.
\item Repaired non-operating channels.
\end{itemize}

\begin {figure}[ht]
\subfigure[]{\includegraphics[width=0.45\textwidth]{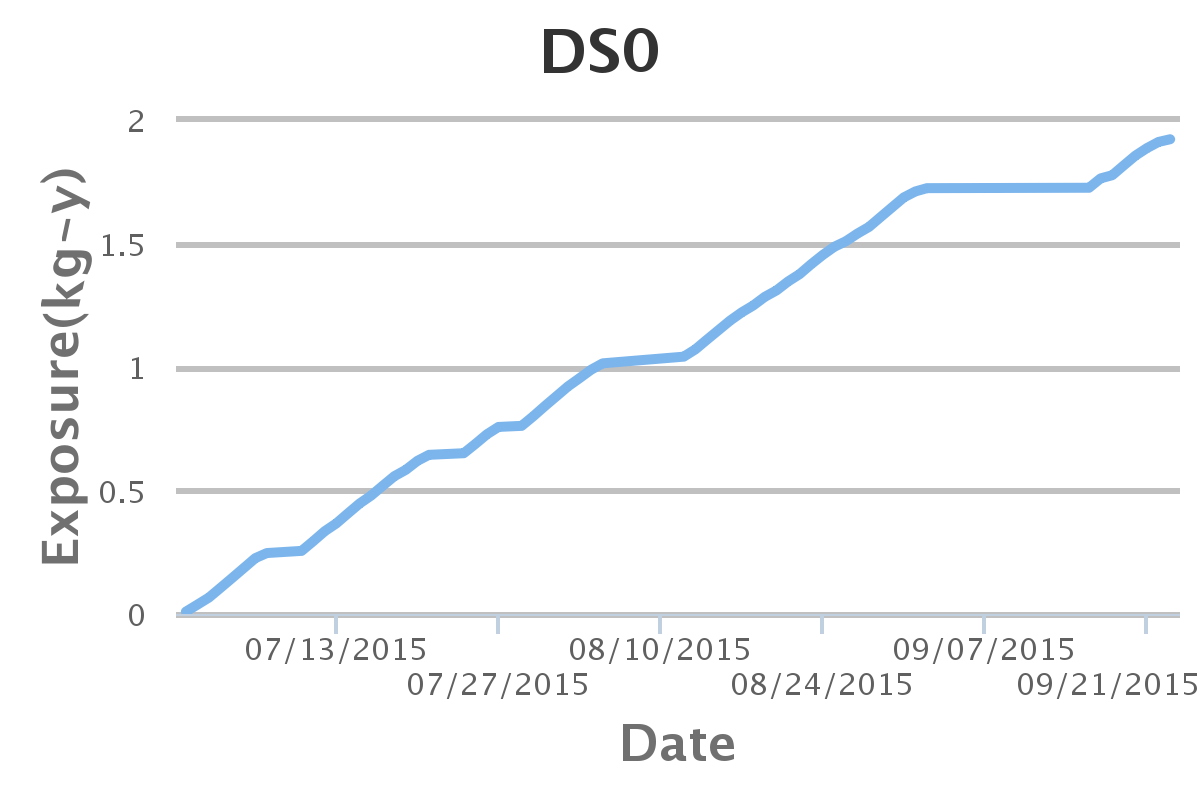}}
\subfigure[]{\includegraphics[width=0.45\textwidth]{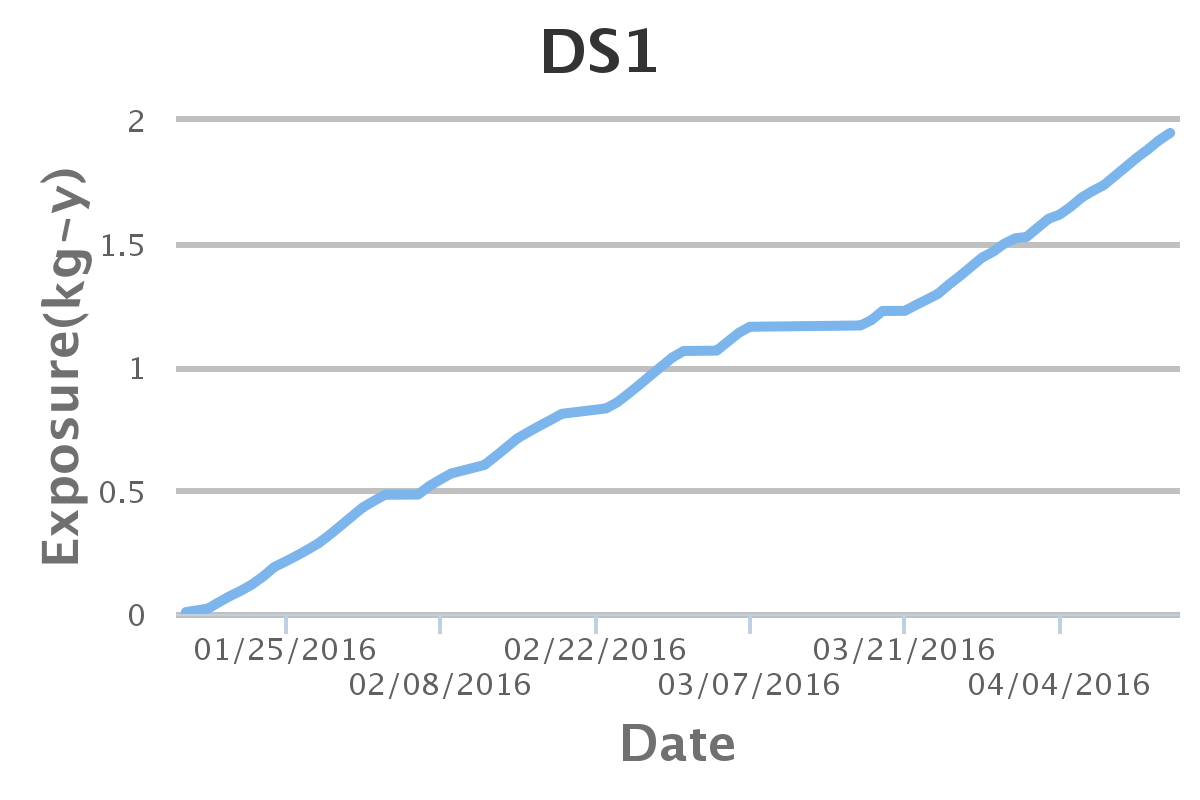}}
\centering \caption{\it DS0 and DS1 exposure plots of the physics data.}
\label{fig:exposure}
\end {figure}

These changes define the difference between DS0 and DS1. Hence, DS1 is the dataset that is being used to determine the background. DS1 data described here are taken from December 2015 to April 2016. Data taking continues, but after that date data blinding began. Table \ref{tab:runtime} summarizes the distribution of the total time elapsed during DS0 and DS1. The exposure evolution, taking into account the total active mass is shown in Figure~\ref{fig:exposure}.

\begin{table}[ht]
\begin{center}
\begin{tabular}{lrr}
\hline\noalign{\smallskip}
& DS0 & DS1\\
\hline\noalign{\smallskip}
Total                               &103.15 d & 104.68 d\\
Total acquired                      &87.93 d &97.52 d\\
Physics                             &47.70 d &54.73 d\\
High radon                          &11.76 d &7.32 d\\
Disruptive commissioning tests      &13.10 d &28.61 d\\
Calibration                         &15.44 d &6.86 d\\
Down time                           &15.21 d & 7.16 d\\
\noalign{\smallskip}\hline
\end{tabular}
\caption{\it DS0 and DS1 duty cycles. The run time is distributed as follows: Total: total - time elapsed from the beginning until the end of the dataset; total acquired - total livetime of the dataset; physics - background runs to be used in the physics analysis; high radon - runs during which the radon purge for the shield was compromised, either intentionally or accidentally; disruptive commissioning tests - data corresponding to DAQ tests and electronics calibrations; calibration - calibration runs taken with a $^{228}$Th or $^{60}$Co source; down time - difference between the total and the acquired time, where the dominant period correspond to detectors being biased down to troubleshoot some detectors.}
\label{tab:runtime}
\end{center}
\end{table}

\begin{figure} [ht]
\begin{center}
 \includegraphics[width=0.7\textwidth]{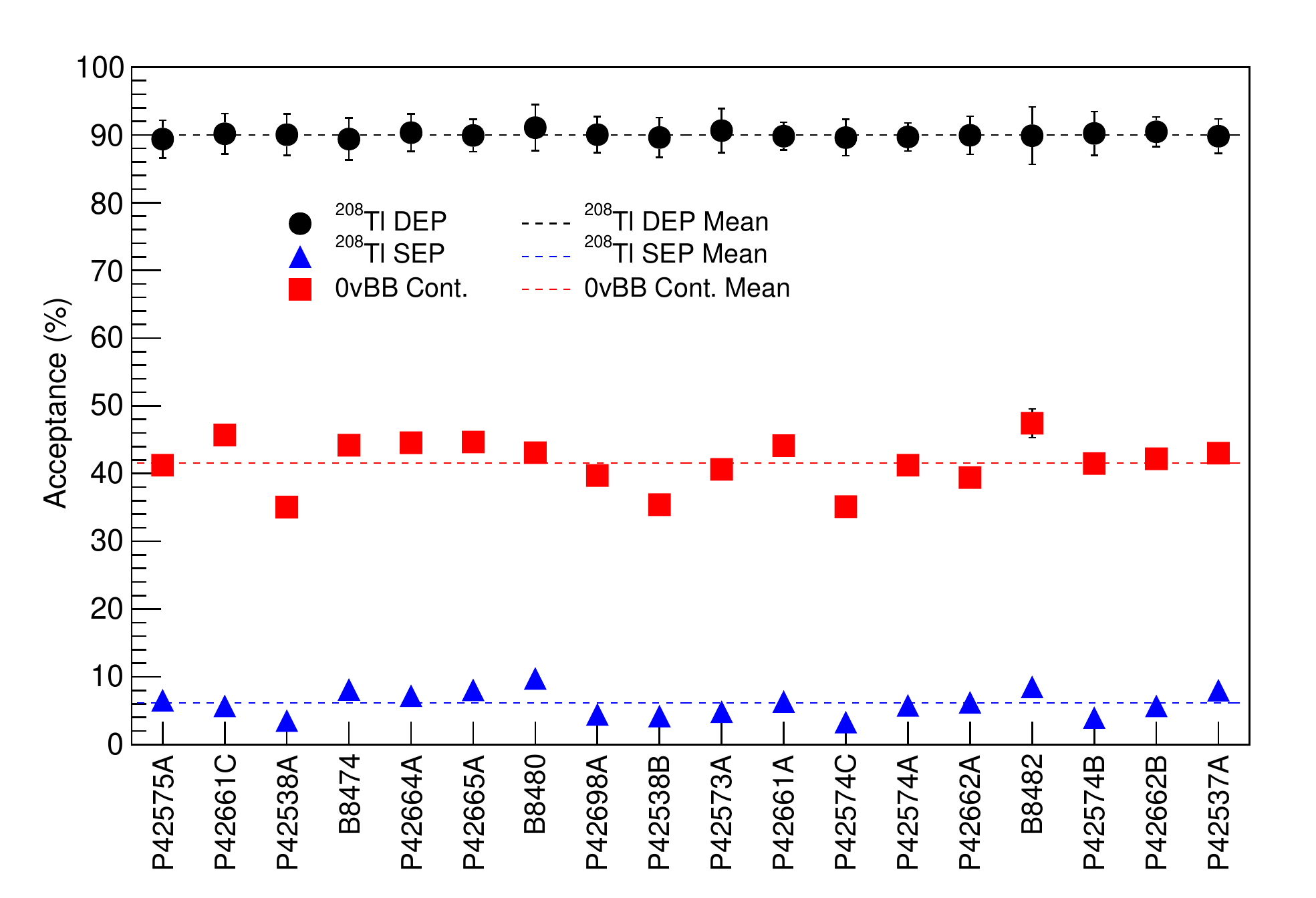}
  \caption{Acceptance efficiency of the pulse shape discrimination cuts at the double-escape peak (single-site events), the single-escape peak (multi-site events), and at \BBz\ ROI (Compton scattering events) of the Module~1 detectors evaluated with $^{228}$Th calibration data.}
  \label{fig:psd}
  \end{center}
\end{figure}

The state-of-the-art data analysis techniques that further enable the \dem $'$s physics reach are still being developed. Double beta decay events are characterized as single-site events because the range of the electron is small compared to that of a typical Compton-scattering background gamma. Using pulse shape discrimination methods~\cite{AEgerda}, it is possible reject $>$90\% of multi-site events while retaining 90\% of single-site events and reducing the Compton continuum at \QBB\ by $>$50\% in case of backgrounds from the $^{228}$Th calibration source, as seen in Figure~\ref{fig:psd}.

\section{Next generation \ge\ experiment}
\label{sec5}

The \mj\ Collaboration is working cooperatively with GERDA and others towards the establishment of a single \ge\ \BBz-decay collaboration to build a large experiment to explore the inverted hierarchy region. Periodic joint meetings take place with this purpose. Leading a tonne-scale \BBz\ experiment is one of the highest priorities new activity for the US Nuclear Physics community as indicated in the latest Long Range Plan for nuclear science in the US recently released by the Nuclear Science Advisory Committee~\cite{LRP}. We anticipate down-select of best technologies, based on results of the two experiments. Moving forward is predicated on demonstration of projected backgrounds. There is on-going work to go from a conceptual design to a viable, competitive proposal including R\&D on: robust signal and high voltage connectors, ultra-clean materials, alternative detector designs, detector signal readout, cryostat and detector mount designs, enrichment, cooling and shielding, and required depth.

\section*{Summary}
\label{sec6}

The~\mj~\dem\ will search for the \BBz-decay with and array of \ge\ detectors. A modular instrument composed of two cryostats built from ultra-pure electroformed copper is being constructed. Module 1 has been taking data since June 2015, and before data blinding started in April 2016, 185 days of data were taken from which 102 days are used in the physics analysis. The data are divided in two datasets: DS0 and DS1. In Fall 2015, we implemented planned improvements: installed the inner copper shield, added additional shielding within the vacuum of the cross arm, exchanged the cryostat seal for one with a low background component, and repaired non-operating channels. These changes define the difference between DS0 and DS1, and DS1 is being used to determine the background. Data analysis is in progress and includes, for instance, pulse shape discrimination to reject multi-site events. The \dem\ aims to reach a background rate of 3~counts/(ROI-t-y) after analysis cuts. This projects to a background level of 1~count/(ROI-t-y) in a large scale experiment that is already being planned.

\section*{Acknowledgments}
\setlength{\emergencystretch}{5em}

This material is based upon work supported by the U.S. Department of Energy, Office of Science, Office of Nuclear Physics under Award  Numbers DE-AC02-05CH11231, DE-AC52-06NA25396, DE-FG02-97ER41041, DE-FG02-97ER41033, DE-FG02-97ER41042, DE-SC0012612, DE-FG02-10ER41715, DE-SC0010254, and DE-FG02-97ER41020. We acknowledge support from the Particle Astrophysics Program and Nuclear Physics Program of the National Science Foundation through grant numbers PHY-0919270, PHY-1003940, 0855314, PHY-1202950, MRI 0923142 and 1003399. We acknowledge support from the Russian Foundation for Basic Research, grant No. 15-02-02919. We  acknowledge the support of the U.S. Department of Energy through the LANL/LDRD Program. This research used resources of the Oak Ridge Leadership Computing Facility, which is a DOE Office of Science User Facility supported under Contract DE-AC05-00OR22725. This research used resources of the National Energy Research Scientific Computing Center, a DOE Office of Science User Facility supported under Contract No. DE-AC02-05CH11231. We thank our hosts and colleagues at the Sanford Underground Research Facility for their support.

\bibliographystyle{elsarticle-num}
\bibliography{skeleton}



\end{document}